\documentclass[epj]{svjour}
\usepackage{graphics,amsmath,amsfonts,color,graphicx}
\def\dd{{\mathrm{d}}}
\def\id{{\rm 1\kern-.25em I}}
\def\s#1{{#1\kern-.5em /}}
\newcommand{\tf}[2]{{\textstyle \frac{#1}{#2}}} 
\newcommand{\ii}{{\textrm{i}}} 
\newcommand{\bv}[1]{{\boldsymbol{#1}}} 
\begin{document}
\title{Static observables of relativistic three-fermion systems with
instantaneous interactions}
\subtitle{\small \textnormal{HISKP-TH-06/02}}
\author{Christian~Haupt, Bernard Metsch \and Herbert-R.~Petry
}                     
%
%
\authorrunning{Christian Haupt et al.}
\titlerunning{Static observables of relativistic three-fermion systems with
instantaneous interactions }
\institute{Helmholtz-Institut f\"ur Strahlen- und Kernphysik,
Nu{\ss}allee 14-16, D-53115 Bonn, Germany, e-mail:
\textrm{haupt@itkp.uni-bonn.de}}
\date{February 16, 2006}
%
\abstract{
We show that static properties like the charge radius and the magnetic moment
of relativistic three-fermion bound states with instantaneous interactions can
be formulated as expectation values with respect to intrinsically defined
wavefunctions. The resulting operators can be given a natural physical
interpretation in accordance with relativistic covariance. We also indicate
how the formalism may be generalized to arbitrary moments. The method is
applied to the computation of static baryon properties with numerical results
for the nucleon charge radii and the baryon octet magnetic
moments. In addition we make predictions for the magnetic moments of some
selected nucleon resonances and discuss the decomposition of the nucleon
magnetic moments in contributions of spin and angular momentum, as well as
the evolution of these contributions with decreasing quark mass.
} 
\maketitle
\section{Introduction}
\label{intro}
Static observables of bound state systems in field theoretic descriptions are
usually extracted from form
factors in the limit of vanishing squared four-momentum transfer of the probing
exchange particle. For example the mean square charge radius is defined as the
slope of the electric form factor at $Q^2 = 0$ and the magnetic moment is
the value of the magnetic form factor at the photon point. Although such an
approach is suitable to produce pure numbers it hardly leads to any insight
into the underlying structure of the observable. It is for example well known
how a fermion produces a magnetic moment through both its spin and its angular
motion, but how does that translate into the magnetic moment of a bound state
e.~g. a baryon composed of three quarks?

On the other hand static properties in non-relativistic quantum mechanics can
be formulated by means of expectation
values involving essentially scalar products of wave functions. The
nonrelativistic
charge radius of a composite system of $N$ particles for example is given
by
\begin{equation}
\langle r^2 \rangle =
\frac{\langle \psi | \sum_{i=1}^N q_i (\bv{x_i} - \bv{R})^2| \psi \rangle}
{Q \langle \psi | \psi \rangle},
\end{equation}
where $q_i$ is the charge of particle $i$, $\bv{x_i}$ its position, $\bv{R}$
the center of mass coordinate and $Q$ the net charge of the system. A direct
relativistic
generalization of this expression is unknown. In this paper we will focus our
attention to a quark model description of baryons. The generalization to other
systems is then quite obvious.
We will show that a synthesis of both approaches mentioned before is indeed
possible -- 
at least if certain restrictions are made to the kind of interactions between
the constituents of the bound system -- and leads to new insights into the
structure of static properties. Moreover the actual computation of static
moments is then easier and also numerically more reliable in comparison to the
computation of form factors in the limit $Q^2 \to 0$.

We work within the framework of the
Bethe-Salpeter equation which has been successfully applied to e.~g. baryon
mass spectra \cite{Loring:2001kv,Loring:2001kx,Loring:2001ky} and form factors
\cite{Merten:2002nz,VanCauteren:2003hn}.

First we briefly outline the Bethe-Salpeter formalism. Details may be found
in ref. \cite{Loring:2001kv}. The construction of current matrix elements from
Bethe-Salpeter amplitudes is addressed, of which a detailed discussion is
given in
\cite{Merten:2002nz}. Since by far the most interesting static observables for
physical applications are mean square charge radii and magnetic moments, we
show how they can be formulated as expectation values with respect to Salpeter
amplitudes. These amplitudes turn out to be the natural quantities which
replace the non-relativistic wavefunctions and possess a canonical scalar
product. Static observables are then represented by certain well defined
operators whose expectation values are computed with the help of this scalar
product.
The emerging operators can be given a natural physical
interpretation and show very interesting structures.
To demonstrate the relevance of this formalism, we apply it to a
concrete physical model for baryons described in
refs. \cite{Loring:2001kv,Loring:2001kx,Loring:2001ky} to compute nucleon
charge radii and magnetic moments and indicate how higher moments can in
principle be computed.
\section{Bethe-Salpeter equation and current matrix elements}
\subsection{Bethe-Salpeter equation}
The basic quantity which describes three fermions as a bound state is the
Bethe-Salpeter amplitude which is defined in position space through:
\begin{equation}
{\chi_{\bar{P}}}_{a_1 a_2 a_3} :=
\langle 0 | T \psi_{a_1}^1(x_1) \psi_{a_2}^2(x_2) \psi_{a_3}^3(x_3)
| \bar{P} \rangle,
\end{equation}
where $\psi_{a_i}^i(x_i)$ are fermion field operators given in the Heisenberg
picture and $a_i$ are multi-indices in Dirac-space and any internal space
which represents a degree of freedom the particle may have. $T$ is
the time ordering operator.
Here $| 0 \rangle$ denotes the physical i.~e. interacting vacuum and
$| \bar{P} \rangle $ denotes a three-fermion bound state with total
four-momentum $\bar{P}$ on the mass shell i.~e. $\bar{P}^2 = M^2$.
Because
of translational invariance it is convenient to introduce a
center-of-mass coordinate $X$ and so-called Jacobi coordinates $\xi$ and
$\eta$:
\begin{align}
X &:= \tf13(x_1 + x_2 + x_3) & x_1 &= X + \tf12\xi + \tf13 \eta
\nonumber \\
\xi &:= x_1 - x_2 & x_2 &= X - \tf12\xi + \tf13 \eta \\
\eta &:= \tf12(x_1 + x_2 - 2 x_3) & x_3 &= X - \tf23 \eta. \nonumber
\end{align}
The corresponding conjugate momenta are then given by the total four-momentum
$P$ and the two relative momenta $p_\xi$ and $p_\eta$:
\begin{align} \label{momenta}
P &:= p_1 + p_2 + p_3 & p_1 &= \tf13 P + p_\xi + \tf12 p_\eta
\nonumber \\
p_\xi &:= \tf12(p_1 - p_2) & p_2 &= \tf13 P - p_\xi + \tf12 p_\eta \\
p_\eta &:= \tf13(p_1 + p_2 - 2p _3) & p_3 &= \tf13 P -p_\eta. \nonumber
\end{align}
Using this set of coordinates the total momentum dependence factorizes and we
may define the Bethe-Salpeter amplitude in momentum space depending only on
the relative momenta:
\begin{multline}
{\chi_{\bar{P}}}_{a_1 a_2 a_3} (x_1, x_2, x_3) =:
e^{-\ii \langle \bar{P}, X \rangle} \int \frac{\dd^4\! p_\xi}{(2 \pi)^4}
\frac{\dd^4\! p_\eta}{(2 \pi)^4}  \\ \times 
e^{- \ii \langle p_\xi, \xi \rangle}
e^{- \ii \langle p_\eta, \eta \rangle}
{\chi_{\bar{P}}}_{a_1 a_2 a_3} (p_\xi, p_\eta).
\end{multline}
One may write the sum of all two-body interactions in the form of a three-body
interaction kernel:
\begin{multline} \label{K^2}
\bar{K}^{(2)}_{a_1 a_2 a_3; a'_1 a'_2 a'_3} :=
{K^{(2)}_{12}}_{a_1 a_2; a'_1 a'_2} (\tf23 \bar{P} + p_\eta, p_\xi, p'_\xi) \\
\times {S_F^3}^{-1}_{a_3 a'_3} (2 \pi)^4 \delta^{(4)}(p_\eta - p'_\eta) \\
+ \textrm{cycl. perm.}.
\end{multline}
Then by introducing the three-particle propagator
\begin{multline} \label{G0}
{G_0}_{a_1 a_2 a_3; a'_1 a'_2 a'_3} (P, p_\xi, p_\eta, p'_\xi, p'_\eta)
:= (2 \pi)^4 \delta^{(4)}(p_\xi - p'_\xi) \\
\times (2 \pi)^4 \delta^{(4)}(p_\eta - p'_\eta)
{S_F^1}_{a_1 a_1'} (\tf13 \bar{P} + p_\xi + \tf12 p_\eta) \\
\times {S_F^2}_{a_2 a_2'} (\tf13 \bar{P} - p_\xi + \tf12 p_\eta)
{S_F^3}_{a_3 a_3'} (\tf13 \bar{P} - p_\eta),
\end{multline}
the Bethe-Salpeter equation can be written in a compact notation
\begin{equation} \label{BSE}
\chi_{\bar{P}}
= - \ii G_0 \left( K^{(3)} + \bar{K}^{(2)} \right) \chi_{\bar{P}},
\end{equation}
where a summation over multi-indices and momentum integrations is tacitly
understood.

The Bethe-Salpeter equation would be incomplete without a prescription of how
to normalize its solutions. Such a prescription can indeed be found (see ref.
\cite{Loring:2001kv}). In a
covariant form and using our compact notation it reads:
\begin{equation} \label{BS-Norm}
- \ii \overline{\chi}_{\bar{P}}
\left[ P^\mu \frac{\partial}{\partial P^\mu} \left(G_0^{-1} + \ii K^{(3)} +
\ii K^{(2)} \right) \right]_{P=\bar{P}} \chi_{\bar{P}} = 2 M^2,
\end{equation}
where $G_0^{-1}$ is the inverse three-particle propagator, i.~e. the inverse
of (\ref{G0}).
Here we introduced the adjoint Bethe-Salpeter amplitude defined as:
\begin{equation}
{\overline{\chi}_{\bar{P}}}_{a_1 a_2 a_3} :=
\langle \bar{P} | T \bar{\psi}_{a_1}^1(x_1) \bar{\psi}_{a_2}^2(x_2)
\bar{\psi}_{a_3}^3(x_3)
| 0 \rangle.
\end{equation}
\subsection{Salpeter equation}
In order to solve the Bethe-Salpeter equation in physical cases relevant for
e.~g. the structure of hadrons, one applies two approximations. First the full
propagators are replaced by the free ones:
\begin{equation}
S_F^i(p_i) = \frac{\ii}{\s{p}_i - m_i + \ii \epsilon}.
\end{equation}
This approximation accounts for self-energy contributions merely by the
introduction of effective fermion masses $m_i$.
Neglecting retardation effects in the interaction kernels leads to the second,
so called
instantaneous approximation. This assumes that there is no dependence of the
interaction kernels
on the relative energies in the rest frame of the composite system:
\begin{align}
K^{(3)}(P, p_\xi, p_\eta, p'_\xi, p'_\eta) \bigg|_{P=(M, \bv{0})}
&= V^{(3)}(\bv{p_\xi}, \bv{p_\eta}, \bv{p'_\xi}, \bv{p'_\eta}) \\[1.5ex]
K^{(2)}(\tf23 P + p_\eta, p_\xi, p'_\xi) \bigg|_{P=(M, \bv{0})}
&= V^{(2)}(\bv{p_\xi}, \bv{p'_\xi}) \\[1.5ex]
\end{align}
These conditions can be formulated in any reference frame, if all momenta are
replaced by:
\begin{equation}
p_\perp := p - \frac{\langle p, P \rangle}{P^2} P.
\end{equation}
This space-like vector is perpendicular to the total four-momentum and in the
rest frame of the system has the desired form $p_\perp = (0, \bv{p})$. Thus
formal covariance of the Bethe Salpeter equation is maintained.

Adopting both approximations, it is possible to integrate out the dependence on
the relative energies in the Bethe-Salpeter equation (\ref{BSE}), thus reducing
the eight-dimensional integral equation to the six-dimensional \\ Salpeter
equation. This
procedure is straightforward if there are no two-body interactions in the
system. The unconnected part of $\bar{K}^{(2)}$ however makes the
reduction more involved. One is then forced to introduce an effective
three-body kernel $V^{\textrm{eff}}_M$ which accounts for the effect of the
two-body interaction approximately (see \cite{Loring:2001kx} for details). The
effective interaction kernel is then
expanded in powers of $K^{(2)}_M + V^{(3)}_R$, where $V^{(3)}_R$ is the
contribution to the three-body interaction
$V^{(3)} = V^{(3)}_\Lambda + V^{(3)}_R$ that couples to mixed
energy components exclusively. $V^{(3)}_\Lambda$ then correspondingly is the
contribution to the three-particle kernel that involves only pure energy
components. Up to lowest order Born approximation
${V^{\textrm{eff}}_M}^{(1)}$ the corresponding Salpeter equation then reads:
\begin{multline}
\Phi^\Lambda_M(\bv{p_\xi}, \bv{p_\eta}) = \\
\left[ \frac{\Lambda^{+++}}{M - \Omega + \ii \epsilon}
+ \frac{\Lambda^{---}}{M + \Omega - \ii \epsilon} \right]
\gamma^0 \otimes \gamma^0 \otimes \gamma^0 \\
\times \int \frac{\dd^3\! p'_\xi}{(2 \pi)^3} \frac{\dd^3\! p'_\eta}{(2 \pi)^3}
V^{(3)}(\bv{p_\xi}, \bv{p_\eta}; \bv{p'_\xi}, \bv{p'_\eta})
\Phi^\Lambda_M(\bv{p'_\xi}, \bv{p'_\eta}) \\
+ \left[ \frac{\Lambda^{+++}}{M - \Omega + \ii \epsilon}
- \frac{\Lambda^{---}}{M + \Omega - \ii \epsilon} \right]
\gamma^0 \otimes \gamma^0 \otimes \id \\
\times \int \frac{\dd^3\! p'_\xi}{(2 \pi)^3}
V^{(2)}(\bv{p_\xi}, \bv{p'_\xi}) \otimes \id\,
\Phi^\Lambda_M(\bv{p'_\xi}, \bv{p_\eta}) \\
+ \textrm{terms with cycl. perm. of two-body force}.
\end{multline}
Here we introduced the short hand notation
$\Lambda^{\pm \pm \pm} := \Lambda_1^\pm(\bv{p_1}) \otimes \Lambda_2^\pm(\bv{p_2}) \otimes \Lambda_3^\pm(\bv{p_3})$, where $\Lambda_i^\pm(\bv{p_i})$ are
projectors onto positive or negative energy respectively and
$\Omega := \omega_1(\bv{p_1}) + \omega_2(\bv{p_2}) + \omega_3(\bv{p_3})$ is
the sum of the relativistic one-particle energies
$\omega_i(\bv{p_i}) = \sqrt{|\bv{p_i}|^2 + m_i^2}$. The Salpeter equation
involves the Salpeter amplitudes, which are projected onto purely positive
and negative energy components respectively:
\begin{multline}
\Phi_M^\Lambda(\bv{p_\xi}, \bv{p_\eta}) :=
\left[ \Lambda^{+++}(\bv{p_\xi}, \bv{p_\eta})
+ \Lambda^{---}(\bv{p_\xi}, \bv{p_\eta}) \right] \\
\times \int \frac{\dd p^0_\xi}{2 \pi} \frac{\dd p^0_\eta}{2 \pi}
\chi_M(p_\xi, p_\eta)
\end{multline}
The full Bethe-Salpeter amplitude, which is needed to calculate
current matrix elements, can be reconstructed from the Salpeter
amplitude in the following way:
\begin{equation}
\chi_{M} =
\left[ G_0 - \ii G_0 \left(V_R^{(3)} + \bar{K}^{(2)}_M -
{V^{\textrm{eff}}_P}^{(1)} \right) G_0 \right]
\Gamma^\Lambda_M,
\end{equation}
where the so called vertex functions $\Gamma^\Lambda_M$ were introduced which
in lowest order in $V^{\textrm{eff}}_M$ are connected to the Salpeter
amplitudes by:
\begin{equation} \label{SA_Vertex}
\Phi_M^\Lambda =
\ii \left[ \frac{\Lambda^{+++}}{M - \Omega} + \frac{\Lambda^{---}}{M + \Omega}
\right] \gamma^0 \otimes \gamma^0 \otimes \gamma^0 \Gamma_M^\Lambda.
\end{equation}
From the normalization condition (\ref{BS-Norm}) a corresponding normalization
for the Salpeter amplitudes can be deduced, which in Born approximation reads:
\begin{multline} \label{NormSA}
\langle \Phi_M^\Lambda | \Phi_M^\Lambda \rangle =
\int \frac{\dd^3\! p^{}_\xi}{(2\pi)^3} \frac{\dd^3\! p^{}_\eta}{(2\pi)^3}\,
{{\Phi^\Lambda_M}^{\! *}}(\vec{p}^{}_\xi,\vec{p}^{}_\eta)
\Phi_{M}^\Lambda(\vec{p}^{}_\xi,\vec{p}^{}_\eta) \\
= 2M.
\end{multline}
Summation over discrete indices is implicitly understood here.
This norm immediately induces a positive definite scalar product:
\begin{equation} \label{SalSP}
\langle \Phi_1 | \Phi_2 \rangle := \\
\int \frac{\dd^3\! p_\xi}{(2\pi)^3} \frac{\dd^3\! p_\eta}{(2\pi)^3}
{\Phi^*_1} (\bv{p_\xi}, \bv{p_\eta})
\Phi_2 (\bv{p_\xi}, \bv{p_\eta}),
\end{equation}
whose existence is of utmost importance since static observables will be
formulated as expectation values with respect to this scalar product as
announced in the introduction.
\subsection{Current matrix elements}
To compute any electromagnetic observable, we need to know the electromagnetic
current $\langle P, \lambda| j^\mu (x) | P', \lambda' \rangle $ between states
with total four-momenta $P'$ and $P$ and helicities $\lambda'$ and $\lambda$
respectively, where $j^\mu(x)$ is the current operator:
\begin{equation}
j^\mu(x) = : \bar{\psi}(x) \hat{q} \gamma^\mu \psi(x) :,
\end{equation}
with the charge operator $\hat{q}$. This current matrix element can be derived
by studying the response of the system in an external electromagnetic field in
first order of the electromagnetic coupling strength \cite{Merten:2002nz}. One
then finds, that the corresponding correlation function
$G^\mu_{P,P'}(p_\xi,p_\eta,p'_\xi,p'_\eta)$ \\
separates at the poles in the total
energy of the system in the following way:
\begin{multline}
G^\mu_{P,P'}(p_\xi,p_\eta,p'_\xi,p'_\eta) = \\
\frac{1}{4 \omega_\bv{P} \omega'_\bv{P'}}
\frac{\chi_P(p_\xi, p_\eta)}{P^0 - \omega_{\bv{P}} + \ii \epsilon}
\langle P | j^\mu(0)| P' \rangle 
\frac{\chi_{P'}(p_\xi, p_\eta)}{{P'}^0 - \omega_{\bv{P'}} + \ii \epsilon} \\
+ \textrm{regular terms for } P^0 \to \omega_{\bv{P}} \textrm{ and }
{P'}^0 \to \omega_{\bv{P'}}.
\end{multline}
The Mandelstam formalism and minimal coupling deliver an independent way to
determine $G^\mu_{P,P'}$. By comparison one then finds the following current
matrix element \cite{Merten:2002nz}:
\begin{multline} \label{CurMa_V2}
\langle P | j^\mu(0) | P' \rangle =
-3 \int\frac{\dd^4\! p_\xi}{(2\pi)^4} \int\frac{\dd^4\! p_\eta}{(2\pi)^4} \\
\times \overline\Gamma_{P}^\Lambda(p_\xi,p_\eta)
S_F^1(p_\xi+\tf12p_\eta) \otimes
S_F^2(-p_\xi+\tf12p_\eta) \\
\otimes S_F^3(P'-p_\eta) \gamma^\mu \hat{q} S_F^3(P-p_\eta)
\Gamma_{P'}^\Lambda(p_\xi,p_\eta),
\end{multline}
where the adjoint vertex function $\overline\Gamma_{M}^\Lambda(p_\xi,p_\eta)$
is related to $\Gamma_{M}^\Lambda(p_\xi,p_\eta)$ through:
\begin{equation} \label{adj_vert}
\overline\Gamma_{M}^\Lambda(p_\xi,p_\eta) =
- {\Gamma_{M}^\Lambda}^\dag(p_\xi,p_\eta)
\gamma^0 \otimes \gamma^0 \otimes \gamma^0.
\end{equation}
Note that in the current matrix element (\ref{CurMa_V2}) the photon couples
to the third fermion exclusively. The couplings to the other fermions have been
accounted for by the factor of $3$. This is possible, since the vertex
functions, which describe a composite fermion system, are totally
antisymmetric.
With the explicit boost prescription of the vertex function:
\begin{equation}
\Gamma_{P}^\Lambda(p_\xi,p_\eta) =
S_{\Lambda_P} \otimes S_{\Lambda_P} \otimes S_{\Lambda_P}
\Gamma_M^\Lambda(\bv{\Lambda^{-1}p_\xi}, \bv{\Lambda^{-1}p_\eta}),
\end{equation}
the time component of the current matrix element in the Breit frame takes the
form:
\begin{multline} \label{CurMa_time}
\langle \mathcal{P}P | j^0(0) | P \rangle =
-3 \int\frac{d^4p_\xi}{(2\pi)^4} \int\frac{d^4p_\eta}{(2\pi)^4}
\overline\Gamma_{M}^{\Lambda}(\bv{p_\xi}, \bv{p_\eta}) \\
\times \left[ S_F^1(p_\xi+\tf12 p_\eta) \otimes
S_F^2(-p_\xi+p_\eta) \otimes S_F^3(M-p_\eta) \right] \\
\times \left[S_{\Lambda_{P}}^2 \otimes S_{\Lambda_{P}}^2 \otimes \id \right]
\left[\id \otimes \id \otimes  \gamma^0 \hat{q}
S_F^3(M-{\Lambda_{P}^{-1}}^2p_\eta) \right] \\
\times \Gamma_{M}^{\Lambda}(\bv{{\Lambda_{P}^{-1}}^2p_\xi},
\bv{{\Lambda_{P}^{-1}}^2p_\eta}),
\end{multline}
where $\mathcal{P}$ is the space inversion operator, i.~e.
$\mathcal{P}(x^0,\bv{x})=(x^0,-\bv{x})$.
The spatial components are accordingly:
\begin{multline} \label{CurMa_space}
\langle \mathcal{P}P | j^i(0) | P \rangle =
-3 \int\frac{d^4p_\xi}{(2\pi)^4} \int\frac{d^4p_\eta}{(2\pi)^4}
\overline\Gamma_{M}^{\Lambda}(\bv{p_\xi}, \bv{p_\eta}) \\
\times \left[ S_F^1(p_\xi+\tf12 p_\eta) \otimes
S_F^2(-p_\xi+p_\eta) \otimes S_F^3(M-p_\eta) \right] \\
\times \left[
S_{\Lambda_{P}}^2 \otimes S_{\Lambda_{P}}^2 \otimes S_{\Lambda_{P}}^2
\right] \\
\times \left[\id \otimes \id \otimes \hat{q}
\left( \gamma^i + [\gamma^i, S_{\Lambda_{P}}]_- \right)
S_F^3(M-{\Lambda_{P}^{-1}}^2p_\eta) \right] \\
\Gamma_{M}^{\Lambda}(\bv{{\Lambda_{P}^{-1}}^2p_\xi},
\bv{{\Lambda_{P}^{-1}}^2p_\eta}).
\end{multline}
Note that in both cases we commuted a triple tensor product of $S_{\Lambda_P}$
past the three fermion propagators and then did an integral transformation to
obtain two successive boosts. This also explains the appearance of the
commutator $[\gamma^i, S_{\Lambda_{P}}]_-$ in the spatial components of the
current matrix element.
\section{The charge radius}
\label{radii}
\subsection{From charge distributions to charge radii}
In the next two section we review the definition and precise computation of a
particular observable namely the charge radius of a composite system in the
framework of quantum field theory. The results are well known but we add them
here for a better understanding.
Given some charge distribution $\rho(\bv{x})$ one defines its mean square
radius by:
\begin{equation} \label{BasicChaRadDef}
\langle r^2 \rangle
= \frac{1}{Q} \int \dd^3\! x\, |\bv{x}|^2 \rho(\bv{x}).
\end{equation}
The radius is normalized by the net charge $Q$, which is simply the integral of
$\rho(\bv{x})$ over the whole space:
\begin{equation}
Q = \int \dd^3\!x\, \rho(\bv{x}).
\end{equation}
However if the charge distribution has no net charge, the normalization $1/Q$
is of course dropped.
If we turn to quantum mechanical systems, the charge distribution is given by
the time component $j^0(x)$ of the four-vector current of the state
$|\psi\rangle$ that describes the system:
\begin{equation} \label{rho_psi}
\rho(\bv{x}) = \frac{\langle \psi | j^0(\bv{x}) | \psi \rangle}
{\langle \psi | \psi \rangle}
\end{equation}
Such a state $|\psi\rangle$ can be represented as a superposition of momentum
eigenstates
\begin{equation} \label{Superpos}
| \psi \rangle =
\int \frac{\dd^3\! P}{\omega_{\bv{P}}} \psi(\bv{P}) | P \rangle.
\end{equation}
$\psi(\bv{P})$ is the wavefunction in momentum space and the states
$|P\rangle$ are normalized according to:
\begin{equation}
\langle P | P' \rangle = 2 \omega_{\bv{P}} (2 \pi)^3
\delta^{(3)}(\bv{P} - \bv{P'}).
\end{equation}
This immediately fixes the normalization of the states $| \psi \rangle$:
\begin{equation}
\langle \psi | \psi \rangle
= 2 (2 \pi)^3 \int \frac{\dd^3\! P}{\omega_{\bv{P}}} \psi^*(\bv{P}) \psi(\bv{P}).
\end{equation}
Let us further investigate the charge distribution by inserting
(\ref{Superpos}) into (\ref{rho_psi}):
\begin{multline} \label{rho_transl}
\rho(\bv{x}) = 
\frac{1}{\langle \psi | \psi \rangle}
\int \frac{\dd^3\! P}{\omega_{\bv{P}}} \int \frac{\dd^3\! P'}{\omega_{\bv{P'}}}
\exp \left( \ii(\bv{P} - \bv{P'}) \cdot \bv{x} \right) \\
\times \psi^*(\bv{P}) \psi(\bv{P'}) \langle P | j^0(0) | P' \rangle.
\end{multline}
We used space translation invariance here to separate the spatial dependence.
As is well known the integral
$\int \dd^3\! x\, \exp(\ii\, \bv{p} \cdot \bv{x})$ is a representation of the
delta distribution. So it follows
\begin{multline} \label{DeltaDist}
\int \dd^3\! x\, |\bv{x}|^2 \exp\left(\ii\, (\bv{P} - \bv{P'}) \cdot \bv{x}\right)
\\
= - \frac{(2 \pi)^3}{4} (\bv{\nabla_{\!P}} - \bv{\nabla_{\!P'}})^2
\,\delta^{(3)}(\bv{P} - \bv{P'}).
\end{multline}
Using (\ref{Superpos}), (\ref{rho_transl}) and (\ref{DeltaDist}) we obtain:
\begin{multline}
\int \dd^3\! x\, |\bv{x}|^2 \langle \psi | j^0(x) | \psi \rangle
= - \frac{(2 \pi)^3}{4} \int \frac{\dd^3\! P}{\omega_{\bv{P}}}
\int \frac{\dd^3\! P'}{\omega_{\bv{P'}}} \\
\times \exp \left( \ii\, (\omega_\bv{P} - \omega_{\bv{P'}}) x^0 \right)
\psi^*(\bv{P}) \psi(\bv{P'}) \\
\times \langle P | j^0(0) | P' \rangle
(\bv{\nabla_{\!P}} - \bv{\nabla_{\!P'}})^2
\delta^{(3)}(\bv{P} - \bv{P'}).
\end{multline}
On the right hand side of this equation we may now integrate by parts twice and
subsequently do one of the two momentum integrations:
\begin{multline} \label{ThreeContrib}
 \int \dd^3\! x\, |\bv{x}|^2 \langle \psi | j^0(x) | \psi \rangle
= - \frac{(2 \pi)^3}{4} \int \dd^3\! P\, \bigg\{ \\
\frac{| \psi(\bv{P}) |^2}{\omega_{\bv{P}}^2}
\left[(\bv{\nabla_{\!P}} - \bv{\nabla_{\!P'}})^2
\langle P | j^0(0) | P' \rangle \right]_{\bv{P'}=\bv{P}} \\
- \left[(\bv{\nabla_{\!P}} - \bv{\nabla_{\!P'}})^2
\frac{\psi^*(\bv{P})}{\omega_{\bv{P}}}
\frac{\psi(\bv{P'})}{\omega_{\bv{P'}}} \right]_{\bv{P'}=\bv{P}}
\langle P | j^0(0) | P \rangle \\
- \bigg[(\bv{\nabla_{\!P}} - \bv{\nabla_{\!P'}})
\frac{\psi^*(\bv{P})}{\omega_{\bv{P}}}
\frac{\psi(\bv{P'})}{\omega_{\bv{P'}}}
\bigg]_{\bv{P'}=\bv{P}} \\
\cdot \left[(\bv{\nabla_{\!P}} - \bv{\nabla_{\!P'}})
\langle P | j^0(0) | P' \rangle\right]_{\bv{P'}=\bv{P}} \bigg\}
\end{multline}
The last of this three terms vanishes, because $\bv{\nabla_{\!P}}$ and
$\bv{\nabla_{\!P'}}$ change sign under space reflection, in other words are
of odd parity. So if we assume that the states $| P \rangle$ have definite
parity then
$\langle P | j^0(0) \bv{\nabla_{\!P}} | P \rangle = 0$.

So far we have considered wave packets that consist of a superposition of
states with different momenta. To obtain states with definite i.~e. sharp
momenta consider first a Gaussian wave packet with a width proportional to
some parameter $\lambda$:
\begin{equation} \label{GaussWP}
\psi(\bv{P}) = \exp(- |\bv{P}|^2/(2 \lambda^2))
\end{equation}
Before we let $\lambda$ go to zero to define a definite momentum state let us
inspect the second term in (\ref{ThreeContrib}). Because for the Gaussian wave
packet from
(\ref{GaussWP}) the wavefunction is real i.~e. $\psi^*(\bv{P}) = \psi(\bv{P})$
we have:
\begin{equation}
\left[(\bv{\nabla_{\!P}} - \bv{\nabla_{\!P'}})^2
\frac{\psi(\bv{P})}{\omega_{\bv{P}}} \frac{\psi(\bv{P'})}{\omega_{\bv{P'}}}
\right]_{\bv{P'}=\bv{P}} = 0.
\end{equation}
Therefore also this term does not contribute to the charge radius and we are
left with the first term in (\ref{ThreeContrib}) only.

Let us now turn to the limit $\lambda \to 0$ again. Since in
this limit $\exp(- |\bv{P}|^2/\lambda^2)$ is another representation of the
delta distribution we find
\begin{equation}
\lim_{\lambda \to 0} \frac{|\psi(\bv{P})|^2}{\omega_{\bv{P}}
\langle \psi(\bv{P}) | \psi(\bv{P}) \rangle}
= \frac{\delta^{(3)}(\bv{P})}{2 (2 \pi)^3}.
\end{equation}
Inserting this together with the first term of (\ref{ThreeContrib}) into the
basic definition of the charge radius (\ref{BasicChaRadDef}) and performing the
final momentum integration yields
\begin{equation}
\langle r^2 \rangle = - \frac{1}{8 M Q}
(\bv{\nabla_{\!P}} - \bv{\nabla_{\!P'}})^2
\langle P | j^0(0) | P' \rangle \bigg|_{\bv{P'}=\bv{P}=0},
\end{equation}
where $M$ is the rest mass of the system. The current matrix element appearing
here is given in the Breit frame if we make the following transformation:
\begin{multline}
(\bv{\nabla_{\!P}} - \bv{\nabla_{\!P'}})^2
\langle P | j^0(0) | P' \rangle \bigg|_{\bv{P'}=\bv{P}=0} \\
= \Delta_{\bv{P}}
\langle \mathcal{P}P | j^0(0) | P \rangle \bigg|_{\bv{P}=0}.
\end{multline}
We then finally end up with the expression:
\begin{equation} \label{ChaRadCurDis}
\langle r^2 \rangle = - \frac{1}{8 M Q}
\Delta_{\bv{P}} \langle \mathcal{P} P | j^0(0) | P \rangle
\bigg|_{\bv{P}=0}.
\end{equation}
\subsection{From form factors to charge radii}
In the last section the derivation of the charge radius operator started from
defining the mean square radius of a charge distribution. As
is well known there is another definition of the charge radius that involves
the electric form factor of the system. There the charge radius is defined as
the
slope of the electric form factor at the photon point. In this subsection we
investigate the interconnection between both definitions and show that they
indeed coincide.

Let us briefly recall some basic definitions in the context of form factors.
From current conservation and Lorentz invariance the electromagnetic vector
current of a spin-$1/2$ state can be parameterized as follows:
\begin{multline} \label{BarCurPar}
\langle P',\lambda' | j^\mu(0) | P,\lambda \rangle
= e \bar{u}_{\lambda'}(P') \\
\times \bigg[ \gamma^\mu \left( F_1(Q^2) + F_2(Q^2) \right)
- \frac{P'^\mu + P^\mu}{2 M} F_2(Q^2) \bigg]
u_\lambda(P).
\end{multline}
$F_1$ and $F_2$ are the Dirac and Pauli form factors respectively. The
Dirac form factor is normalized to the charge $Q$ whereas the Pauli form
factor is normalized to the anomalous magnetic moment $\kappa$.
Both form factors are functions of the squared invariant momentum transfer
$Q^2 := -q^2 = - (P' - P)^2$. The Dirac spinors are normalized in a Lorentz
invariant fashion:
\begin{equation}
\bar{u}_{\lambda'}(P) u_\lambda(P)
= 2 M\,\delta_{\lambda' \lambda}.
\end{equation}
Using this normalization one shows that
\begin{equation}
\bar{u}_{\lambda'}(\mathcal{P}P) u_\lambda(P)
= 2 \sqrt{M^2 + Q^2/4} \delta_{\lambda' \lambda}
\end{equation}
as well as
\begin{equation}
\bar{u}_{\lambda'}(\mathcal{P}P) \gamma^0 u_\lambda(P)
= 2 M \delta_{\lambda' \lambda}.
\end{equation}
Using both expressions one can write the time component of the electromagnetic
vector current (\ref{BarCurPar}) in the Breit frame as
\begin{equation} \label{BarCurTime}
\langle \mathcal{P}P,\lambda | j^0(0) | P,\lambda \rangle
= 2 e M G_E(Q^2),
\end{equation}
where $G_E(Q^2)$ is the electric Sachs form factor. It is defined together
with the magnetic Sachs form factor as a combination of the Dirac and Pauli
form factors:
\begin{eqnarray} \label{SachsFF}
G_E(Q^2) &:=& F_1(Q^2) - \frac{Q^2}{4 M^2} F_2(Q^2) \\
G_M(Q^2) &:=& F_1(Q^2) + F_2(Q^2).
\end{eqnarray}
The mean square charge radius is defined as the slope of the electric form
factor at the photon point i.~e. at $Q^2 = 0$:
\begin{equation}
\langle r^2 \rangle =
 - \frac{6}{G_E(0)} \frac{\dd G_E(Q^2)}{\dd Q^2} \bigg|_{Q^2 = 0}.
\end{equation}
Since $G_E(0) = F_1(0) = Q$ the normalization $1/G_E(0)$ is dropped
in case the net charge vanishes. From this definition together with
(\ref{BarCurTime}) we then find
\begin{equation} \label{CR_CurMa}
\langle r^2 \rangle =
 - \frac{3}{M Q} \frac{\dd}{\dd Q^2}
\langle \mathcal{P}P,\lambda | j^0(0) | P,\lambda \rangle \bigg|_{Q^2 = 0}.
\end{equation}
This result has to be compared to the one that we obtained in the previous
section, namely equation (\ref{ChaRadCurDis}). There we found the Laplace
operator with respect to $\bv{P}$ instead of a
single derivative with respect to $Q^2$ acting on the current matrix element.
However both expressions turn out to be exactly equal: From the
parameterization
of the vector current (\ref{BarCurPar}) it is clear that the current matrix
element $\langle \mathcal{P}P,\lambda | j^0(0) | P,\lambda \rangle$
depends on $Q^2$. In the Breit frame the dependence of $Q^2$ on the momenta of
the incoming and outgoing bound states becomes rather simple. It reads
$Q^2 = 4 |\bv{P}|^2$. Then for any function $f$ depending on $4 |\bv{P}|^2$ the
following identity holds:
\begin{equation}
\Delta_{\bv{P}} f(4 |\bv{P}|^2) =
4 \left(\Delta_{\bv{P}} |\bv{P}|^2\right) \frac{\dd}{\dd Q^2} f(Q^2) =
24 \frac{\dd}{\dd Q^2} f(Q^2).
\end{equation}
Inserting this into (\ref{ChaRadCurDis}) we see that it coincides with
(\ref{CR_CurMa}). Thus the definition of the charge radius from form
factors is exactly equivalent to that from charge distributions. It must be
noted however that expression (\ref{ChaRadCurDis}) is somewhat more general
than (\ref{CR_CurMa}) in the sense that it is valid for particles with
arbitrary spin. Nevertheless one can show that both expressions lead to the
same result. We decided however to start from (\ref{CR_CurMa}) simply because
it contains only a first order derivative.
\subsection{The charge radius as an expectation value with respect to Salpeter
amplitudes}
We now want to analyze the charge radius in the Bethe-Salpeter framework for a
three-quark system. It is essential to know the
$Q^2$-dependence of the current matrix element (\ref{CurMa_time}).
So let us inspect its $P$-dependent part alone:
\begin{multline}
\left[ S_{\Lambda_{P}}^2 \otimes S_{\Lambda_{P}}^2 \otimes
S_F^3(M-{\Lambda_{P}^{-1}}^2p_\eta) \right]\\
\times \Gamma_{M}^{\Lambda}(\overrightarrow{{\Lambda_{P}^{-1}}^2p_\xi},
\overrightarrow{{\Lambda_{P}^{-1}}^2p_\eta})\\
:= \left[ S_{\Lambda_{P}}^2 \otimes S_{\Lambda_{P}}^2 \otimes \id \right]
f({\Lambda_{P}^{-1}}^2 p_\xi, {\Lambda_{P}^{-1}}^2 p_\eta).
\end{multline}
We now exploit an important property of Lie groups, \\ namely that every group
element may be represented as an exponential mapping of the Lie algebra:
\begin{multline} \label{Boost}
\left[S_{\Lambda_{P}}^2 \otimes S_{\Lambda_{P}}^2 \otimes \id \right]
f({\Lambda_{P}^{-1}}^2p_\xi , {\Lambda_{P}^{-1}}^2p_\eta) \\
= \exp(- 2 \ii\, \bv{\eta}(P) \cdot \bv{\hat{K}}) f(p_\xi,p_\eta)
\end{multline}
The parameter $\bv{\eta}$, commonly called rapidity, is defined as follows:
\begin{equation} \label{rapidity}
\bv{\eta}(P) := \frac{\bv{P}}{P^0} = \frac{-\bv{q}}{2 \sqrt{M^2 + Q^2/4}},
\end{equation}
where the last equality follows from Breit frame kinematics.
The operator $\bv{\hat{K}}$ is an infinitesimal boost. The
generators of the Lorentz group are given by the following skew symmetric
tensors (see e.~g. \cite{Peskin/Schroeder}):
\begin{eqnarray}
J^{\mu\nu} &=& \ii (x^\mu \partial^\nu - x^\nu \partial^\mu) \\[1.5ex]
S^{\mu\nu} &=& \tf\ii4 [\gamma^\mu , \gamma^\nu]^{}_{-},
\end{eqnarray}
Because of skewness there are six independent quantities. $J^{0i}$ are
the three generators of boosts and the remaining three operators generate
rotations (in fact they are the angular momentum operators).
In momentum space we have
$J^{\mu\nu}_p = \ii (p^\mu \partial / \partial p_\nu -
p^\nu \partial / \partial p_\mu)$.
$S^{\mu\nu}$ are the corresponding generators in Dirac space. The infinitesimal
boost $\bv{\hat{K}}$ then simply reads
\begin{equation} \label{Inf_boost}
\begin{split}
\hat{K}^i =&\, -J^{0i}_{p_\xi} - J^{0i}_{p_\eta}
+ S^{0i} \otimes \id \otimes \id + \id \otimes S^{0i} \otimes \id \\[1.5ex]
=&\, \ii \big(
- p_\xi^0 \tf{\partial}{\partial p_\xi^i}
- p_\xi^i \tf{\partial}{\partial p_\xi^0}
- p_\eta^0 \tf{\partial}{\partial p_\eta^i}
- p_\eta^i \tf{\partial}{\partial p_\eta^0} \\
& + \tf12 \alpha^i \otimes \id \otimes \id
+ \id \otimes \tf12 \alpha^i \otimes \id \big).
\end{split}
\end{equation}
Inserting (\ref{Boost}) back into the current matrix element (\ref{CurMa_time})
we find
\begin{multline} \label{CurMa_time_exp}
\langle \mathcal{P}P | j^0(0) | P \rangle =
-3 \int\frac{d^4p_\xi}{(2\pi)^4} \int\frac{d^4p_\eta}{(2\pi)^4}
\overline\Gamma_{M}^{\Lambda}(\bv{p_\xi}, \bv{p_\eta}) \\
\times \left[ S_F^1(p_\xi+\tf12 p_\eta) \otimes
S_F^2(-p_\xi+p_\eta) \otimes S_F^3(M-p_\eta) \right] \\
\times \exp( - 2 \ii\, \bv{\eta}(P) \cdot \bv{\hat{K}})
\left[\id \otimes \id \otimes  \gamma^0 \hat{q}
S_F^3(M-p_\eta) \right] \\
\times \Gamma_{M}^{\Lambda}(p_\xi, p_\eta).
\end{multline}
Since the charge radius is proportional to the slope of the current matrix
element (\ref{CurMa_time_exp}) at $Q^2 = 0$ we are interested in the term of
the expansion linear in $Q^2$ and thus -- because $|\bv{\eta}(P)|^2$ is of
order
$Q^2$ -- linear in $|\bv{\eta}(P)|^2$. Writing the expansion explicitly out up
to this order we have:
\begin{multline}
\langle \mathcal{P}P | j^0(0) | P \rangle =
2 M Q \\
-3 \int\frac{d^4p_\xi}{(2\pi)^4} \int\frac{d^4p_\eta}{(2\pi)^4}
\overline\Gamma_{M}^{\Lambda}(\bv{p_\xi}, \bv{p_\eta}) \\
\times \left[ S_F^1(p_\xi+\tf12 p_\eta) \otimes
S_F^2(-p_\xi+p_\eta) \otimes S_F^3(M-p_\eta) \right] \\
\times
\left[ - 2 \sum_{i,j=1}^3 \eta^i(P) \eta^j(P) \hat{K}^i \hat{K}^j \right]
\\
\times \left[\id \otimes \id \otimes  \gamma^0 \hat{q}
S_F^3(M-p_\eta) \right]
\Gamma_{M}^{\Lambda}(p_\xi, p_\eta)
+ \mathcal{O}(\bv{\eta}^4).
\end{multline}
By inserting this into (\ref{CR_CurMa}), the charge radius then takes the form:
\begin{multline} \label{CR_before_int}
\langle r^2 \rangle =
 - \frac{18}{M Q} \sum_{i,j=1}^3 \left[ \frac{\dd}{\dd Q^2}
\eta^i(P) \eta^j(P) \right]_{Q^2=0} \\
\times \int\frac{d^4p_\xi}{(2\pi)^4} \int\frac{d^4p_\eta}{(2\pi)^4}
\overline\Gamma_{M}^{\Lambda}(\bv{p_\xi}, \bv{p_\eta}) \\
\times \left[ S_F^1(p_\xi+\tf12 p_\eta) \otimes
S_F^2(-p_\xi+p_\eta) \otimes S_F^3(M-p_\eta) \right]
\hat{K}^i \hat{K}^j
\\
\times \left[\id \otimes \id \otimes  \gamma^0 \hat{q}
S_F^3(M-p_\eta) \right]
\Gamma_{M}^{\Lambda}(p_\xi, p_\eta).
\end{multline}
The integration over the relative energies can now be performed by using the
partial fractions decomposition of the fermion propagators:
\begin{equation} \label{PFD_prop}
S_F^i (p_i) = \ii \left(
\frac{\Lambda_i^+(\bv{p_i})}{p_i^0 - \omega_i(\bv{p_i}) + \ii \epsilon}
+ \frac{\Lambda_i^-(\bv{p_i})}{p_i^0 + \omega_i(\bv{p_i}) - \ii \epsilon}
\right) \gamma^0.
\end{equation}
With the aid of Cauchys theorem both integrations can be performed and one
obtains:
\begin{multline} \label{r^2_expl}
\langle r^2 \rangle =
\frac{18}{M Q} \sum_{i,j=1}^3 \left[ \frac{\dd}{\dd Q^2}
\eta^i(P) \eta^j(P) \right]_{Q^2=0} \\
\int\frac{\dd^3p_\xi}{(2\pi)^3} \int\frac{\dd^3p_\eta}{(2\pi)^3}
\overline\Gamma_{M}^{\Lambda}(\boldsymbol{p_\xi},\boldsymbol{p_\eta})
\left[ \frac{\Lambda^{+++}}{(M-\Omega)}
+ \frac{\Lambda^{---}}{(M+\Omega)} \right] \\
\times \hat{K}'^i \hat{K}'^j \hat{q}^3
\left[
\frac{\Lambda^{+++}}{(M-\Omega)}
+ \frac{\Lambda^{---}}{(M+\Omega)} \right] \\
\times [\gamma^0 \otimes \gamma^0 \otimes \gamma^0]
\Gamma_{M}^{\Lambda}(\boldsymbol{p_\xi},\boldsymbol{p_\eta}).
\end{multline}
Now $\hat{q}^3$ denotes the charge operator acting on the third fermion.
After integration the boost becomes:
\begin{equation} \label{K'^i}
\hat{K}'^i := -\tf12 (\omega_1 - \omega_2) \ii \tf{\partial}{\partial p_\xi^i}
- (\omega_1 + \omega_2) \ii \tf{\partial}{\partial p_\eta^i}
- \tf{\ii p_1^i}{2 \omega_1} - \tf{\ii p_2^i}{2 \omega_2}.
\end{equation}
Note that we used the anticommutator $\{\gamma^0,\alpha^i\}_+=0$ and the
relation
$\Lambda_i^\pm \alpha^j = \alpha^j \Lambda_i^\mp \pm p_i^j/\omega_i$ here.
Now by using relations (\ref{SA_Vertex}) and (\ref{adj_vert}) one can replace
the vertex functions in (\ref{r^2_expl}) by Salpeter amplitudes. The result
is an expectation value with respect to the Salpeter scalar product
(\ref{SalSP}):
\begin{multline} \label{r^2_comp}
\langle r^2 \rangle =
\frac{18}{M Q} \sum_{i,j=1}^3 \left[ \frac{\dd}{\dd Q^2}
\eta^i(P) \eta^j(P) \right]_{Q^2=0} \\
\times \langle \Phi_M^\Lambda | \hat{K}'^i \hat{K}'^j \hat{q}^3
| \Phi_M^\Lambda \rangle,
\end{multline}
Since $\bv{\hat{K}'}$ is a tensor operator of rank $1$, that is a vector
operator,
$\hat{K}'^i \hat{K}'^j$ is a Cartesian tensor operator of rank $2$. As is
well known, every Cartesian tensor may be decomposed into irreducible
representations of the rotation group $SO(3)$. The decomposition of a rank $2$
tensor $T_{ij}$ is given by:
\begin{multline} \label{Tens_Decomp}
T_{ij} = \tf13 \textrm{tr}(T) \delta_{ij} + \tf12 (T_{ij} - T_{ji}) \\
+ \tf12 (T_{ij} + T_{ji} - \tf23 \textrm{tr}(T) \delta_{ij}).
\end{multline}
According to their transformation properties under rotations, the first term
belongs to the scalar representation, the second to the vector representation
and the last to the five dimensional representation of spin $2$. Let us now
address the question, which of these representations will vanish due to
selection rules in the scalar product in (\ref{r^2_comp}). Let us start with
the vector representation:
\begin{multline}
\sum_{i,j=1}^3 \left[ \frac{\dd}{\dd Q^2}
\eta^i(P) \eta^j(P) \right]_{Q^2=0} \\
\times \langle \Phi_M^\Lambda |
\tf12 \left( \hat{K}'^i \hat{K}'^j - \hat{K}'^j \hat{K}'^i \right) \hat{q}^3
| \Phi_M^\Lambda \rangle = 0.
\end{multline}
This is so, because $\eta^i(P) \eta^j(P)$ is symmetric, whereas \\
$\hat{K}'^i \hat{K}'^j - \hat{K}'^j \hat{K}'^i$ is antisymmetric under the
exchange of indices. For the spin $2$ representation we cite the Wigner-Eckart
theorem and in particular the triangularity relation which states, that for
spherical tensor operators of rank $k$:
\begin{equation}
\langle j_1 | T^{[k]}_q | j_2 \rangle = 0\quad \textrm{unless}\quad
|j_1 - j_2| \le k \le j_1 + j_2.
\end{equation}
In our case $j_1 = j_2 = \tf12$ and $k=2$,
so the spin $2$ representation in (\ref{Tens_Decomp}) gives no contribution.
Only the scalar representation contributes and we get from (\ref{Tens_Decomp})
and (\ref{r^2_comp}):
\begin{equation}
\langle r^2 \rangle =
\frac{6}{M Q} \left[ \frac{\dd}{\dd Q^2} \bv{\eta}^2(P) \right]_{Q^2=0}
\langle \Phi_M^\Lambda | \bv{\hat{K'}}^2 \hat{q}^3 | \Phi_M^\Lambda \rangle.
\end{equation}
Recalling the definition of the rapidity (\ref{rapidity}) we find:
\begin{equation} \label{der_rap}
\frac{\dd}{\dd Q^2} \bv{\eta}^2(P) \bigg|_{Q^2=0}
= \frac{1}{4 M^2},
\end{equation}
which brings us almost to our final result:
\begin{equation} \label{CharRad_K}
\langle r^2 \rangle =
\frac{3}{2 M Q}
\langle \Phi_M^\Lambda |
\frac{\bv{\hat{K'}}^2}{M^2} \hat{q}^3
| \Phi_M^\Lambda \rangle,
\end{equation}
By rewriting $\hat{K'}^i$ in terms of one-particle coordinates we find:
\begin{equation}
\hat{K}'^i = \frac12 \left[ \Omega \left(\ii \tf{\partial}{\partial p_3^i}
- \frac1\Omega \sum_{\alpha=1}^3 \omega_\alpha \ii
\tf{\partial}{\partial p_\alpha^i} \right) + \textrm{h.~c.} \right].
\end{equation}
It is also useful to define:
\begin{equation} \label{Def_R}
\bv{\hat{R}} := \frac1\Omega \sum_{\alpha=1}^3 \omega_\alpha \ii
\bv{\nabla}_{p_\alpha}.
\end{equation}
The expression (\ref{CharRad_K}) is still not symmetric in all three particles.
The third fermion
seems to play a special role. However this asymmetry is only due to the fact
that in deriving the current matrix element we exploited
the total asymmetry of the vertex functions under particle interchange and
coupled the photon to the third fermion exclusively accounting for the other
couplings by multiplying with a factor of $3$ (see \cite{Merten:2002nz}). If we
reverse this procedure,
cancel the factor of $3$ and symmetrize the expression over the three
particles, we end up with a symmetric version:
\begin{multline} \label{CharRad_boost_final}
\langle r^2 \rangle =
\frac{1}{Q\, \langle \Phi^\Lambda_M | \Phi^\Lambda_M \rangle} \\
\times \langle \Phi^\Lambda_M |
\sum_{\alpha=1}^3 \frac{\hat{q}^\alpha}{4}
\left[ \frac{\Omega}{M} \left(
\ii \bv{\nabla_{p_\alpha}} - \bv{\hat{R}} \right) + \textrm{h.~c.}
\right]^2 | \Phi^\Lambda_M \rangle.
\end{multline}
Now the sum runs over all three fermions.
Note that we also used the norm of the Salpeter amplitudes (\ref{NormSA}) to
replace $2M$ by
$\langle \Phi^\Lambda_M | \Phi^\Lambda_M \rangle$.
\subsection{Interpretation}
\begin{figure}
\begin{center}
\includegraphics[bb=66 189 595 841,clip,angle=90,width=\linewidth]{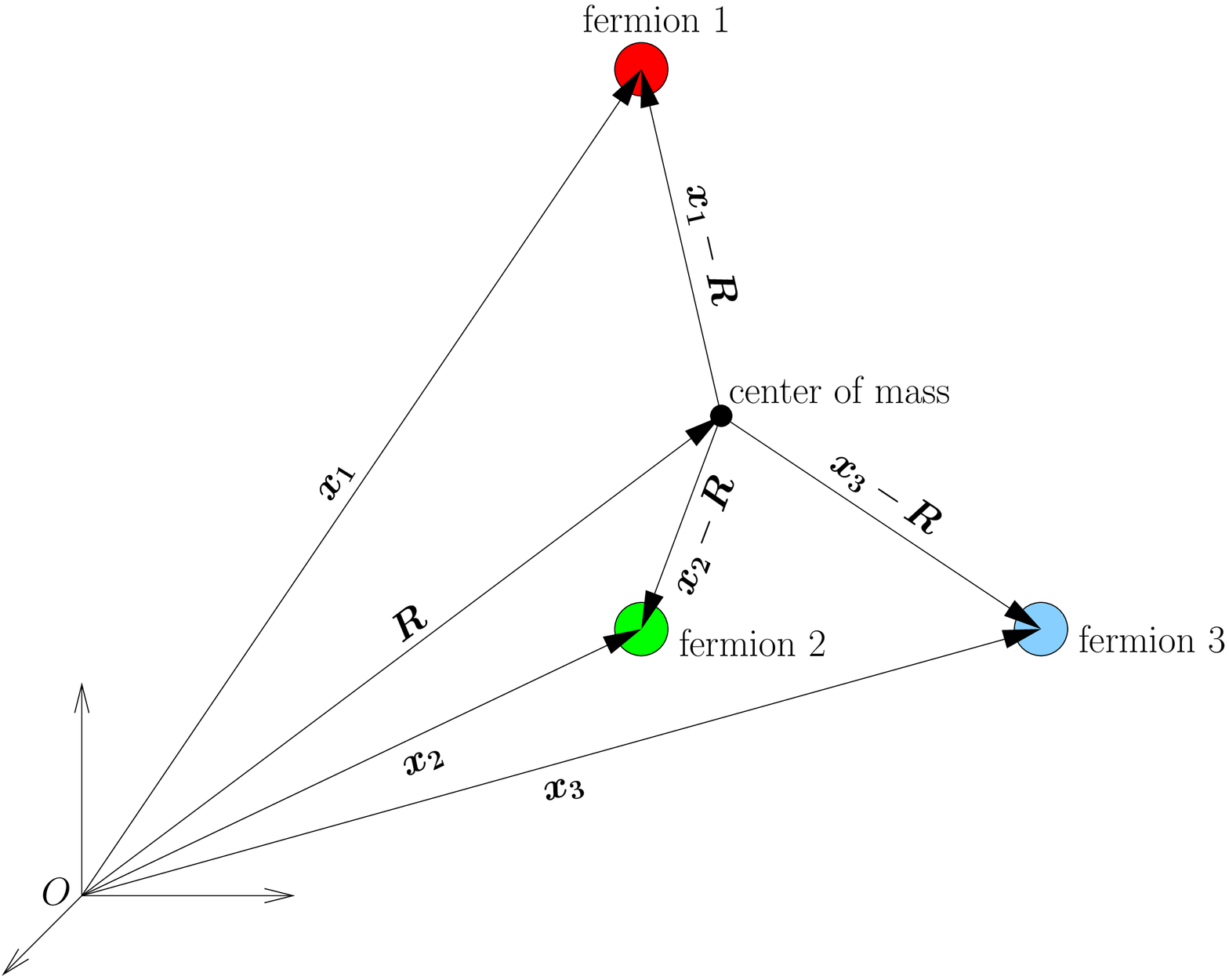}
\end{center}
\caption{$O$ denotes the origin of the reference frame, $\bv{R}$ the position
vector of the relativistic center of mass and $\bv{x_1}$ through $\bv{x_3}$ the
position vectors of the three fermions.}
\label{Fig_CMS}
\end{figure}

Having derived an analytic expression for the mean square charge radius of a
bound three-fermion system with instantaneous interaction kernels,
it is worthwhile to give the result a meaningful physical interpretation.
To interpret the operator between the Salpeter amplitudes it is useful to note
that
$\ii \bv{\nabla_{p_\alpha}}$ is the position operator in momentum space:
\begin{equation}
\ii \bv{\nabla_{p_\alpha}} \equiv \bv{\hat{x}_{\alpha}}.
\end{equation}
Consequently the quantity $\bv{\hat{R}}$ as defined in (\ref{Def_R}) is the
canonical relativistic center of mass of a three-particle system:
\begin{equation}
\bv{\hat{R}} = \frac1\Omega \sum_{\alpha=1}^3 \omega_\alpha
\bv{\hat{x}_{\alpha}}.
\end{equation}
At fermion momenta small compared to their masses, i.~e.
$|\bv{p_\alpha}| \ll m_\alpha$, we have $\omega_\alpha \to m_\alpha$ and thus
the expression reduces to the well known nonrelativistic center of mass:
\begin{equation}
\bv{\hat{R}}_{\textrm{nr}} =
\frac{1}{m_1 + m_2 + m_3} \sum_{\alpha=1}^3 m_\alpha
\bv{\hat{x}_{\alpha}}.
\end{equation}
The expression
\begin{equation} \label{pos_cms}
\ii \bv{\nabla_{p_\alpha}} - \bv{\hat{R}} =
\bv{\hat{x}_{\alpha}} - \frac1\Omega \sum_{\beta=1}^3 \omega_\beta
\bv{\hat{x}_{\beta}}
\end{equation}
then corresponds to the position of particle $\alpha$ as measured from the
relativistic center
of mass. Fig. \ref{Fig_CMS} illustrates the situation.
Since $\ii \bv{\nabla_{p_\alpha}} - \bv{\hat{R}}$ is the difference between two
vector operators, it is invariant under translations and consequently the mean
square radius (\ref{CharRad_boost_final}) is translationally invariant.
Finally we want to call attention to the relativistic factor $\Omega/M$
in (\ref{CharRad_boost_final}) which weights the relative distance of each
fermion with the collective relativistic energy. Therefore an enhancement can
be seen the more relativistic the system is.
\section{The magnetic moment}
\subsection{The magnetic moment as an expectation value}
To derive magnetic moments from form factors in an analogous way we start from
the parameterization
(\ref{BarCurPar}) of the electromagnetic vector current:
\begin{multline} \label{BarCurPlus}
\langle \mathcal{P}P,\lambda' | j^+(0) | P,\lambda \rangle \\
= e  \left[ F_1(Q^2) + F_2(Q^2) \right]
\bar{u}_{\lambda'}(\mathcal{P}P) \gamma^+ u_\lambda(P),
\end{multline}
where the ``+''-component of the current is defined by:
\begin{equation} \label{j_+}
j^+(0) = j^1(0) + \ii j^2(0).
\end{equation}
Note that this definition deviates from the definition of the components of a
spherical tensor operator of rank 1 which are usually given by:
\begin{equation} \label{vector_operator}
T^{[1]}_\pm := \mp \frac{1}{\sqrt{2}} (T_1 \pm \ii T_2)\quad \textrm{and}\quad
T^{[1]}_0 := T_3.
\end{equation}
The total spin of the system makes a spin flip of
course so we have $\lambda' \ne \lambda$. The spin polarizations will be fixed
later. Evaluation of the spinorial part of this
equation yields:
\begin{equation}
\bar{u}_{\lambda'}(\mathcal{P}P) \gamma^+ u_\lambda(P) = 2 \sqrt{Q^2}.
\end{equation}
Together with the definition of the magnetic Sachs form factor (\ref{SachsFF})
we then get from (\ref{BarCurPlus}) the relation:
\begin{equation}
G_M(Q^2) \label{GM}
= \frac{\langle \mathcal{P}P,\lambda' | j^+(0) | P,\lambda \rangle}
{2 \sqrt{Q^2}},
\end{equation}
which expresses the magnetic form factor in terms of spatial components of
the current matrix element. The magnetic moment is defined
as the value of the magnetic form factor at the photon point:
\begin{equation} \label{Def_MaMo_FF}
\langle \mu \rangle := G_M(Q^2 = 0).
\end{equation}
Because of the denominator in (\ref{GM}) taking this limit requires some care.
We need to know the $Q^2$-dependence of the current matrix element. But
first let us choose the three-momentum transfer to point in the $3$-direction
from now on:
\begin{equation} \label{Def_q_z}
\bv{q} := \left( \begin{array}{c} 0 \\ 0 \\ q_3 \end{array} \right)
= \sqrt{Q^2}\, \bv{e_3}.
\end{equation}
With this choice we have $[\gamma^+, S_{\Lambda_{P}}]_-=0$ such that the
current matrix element (\ref{CurMa_space}) now becomes:
\begin{multline} \label{CurMa_space_q3}
\langle \mathcal{P}P | j^+(0) | P \rangle =
-3 \int\frac{d^4p_\xi}{(2\pi)^4} \int\frac{d^4p_\eta}{(2\pi)^4}
\overline\Gamma_{M}^{\Lambda}(\bv{p_\xi}, \bv{p_\eta}) \\
\times \left[ S_F^1(p_\xi+\tf12 p_\eta) \otimes
S_F^2(-p_\xi+p_\eta) \otimes S_F^3(M-p_\eta) \right] \\
\times \left[
S_{\Lambda_{P}}^2 \otimes S_{\Lambda_{P}}^2 \otimes S_{\Lambda_{P}}^2
\right]
\left[\id \otimes \id \otimes \hat{q} \gamma^+
S_F^3(M-{\Lambda_{P}^{-1}}^2p_\eta) \right] \\
\Gamma_{M}^{\Lambda}(\bv{{\Lambda_{P}^{-1}}^2p_\xi},
\bv{{\Lambda_{P}^{-1}}^2p_\eta})
\end{multline}
To extract the $Q^2$-dependence of this expression we express the boost as an
exponential just like in the previous section on charge radii:
\begin{multline} \label{Expand_CurMa_space}
\langle \mathcal{P}P, \lambda' | j^+(0) | P, \lambda \rangle = 
-3 \int\frac{d^4p_\xi}{(2\pi)^4} \int\frac{d^4p_\eta}{(2\pi)^4} \\
\times \overline\Gamma_{M,\lambda'}^{\Lambda}(\bv{p_\xi}, \bv{p_\eta}) \\
\times \left[ S_F^1(p_\xi+\tf12 p_\eta) \otimes
S_F^2(-p_\xi+p_\eta) \otimes S_F^3(M-p_\eta) \right] \\
\times \exp(- 2 \ii\, \bv{\eta}(P) \cdot \bv{\hat{K}})
\left[\id \otimes \id \otimes \hat{q} \gamma^+
S_F^3(M-p_\eta) \right] \\
\times \Gamma_{M,\lambda}^{\Lambda}(p_\xi,p_\eta).
\end{multline}
This time however the boost generator also acts on the Dirac space of the third
fermion as can be seen from the current matrix element (\ref{CurMa_space_q3}):
\begin{multline}
\hat{K}^i = \ii \bigg[
- p_\xi^0 \tf{\partial}{\partial p_\xi^i}
- p_\xi^i \tf{\partial}{\partial p_\xi^0}
- p_\eta^0 \tf{\partial}{\partial p_\eta^i}
- p_\eta^i \tf{\partial}{\partial p_\eta^0} \\
+ \tf12 \left(\alpha^i \otimes \id \otimes \id
+ \textrm{cycl. perm.} \right) \bigg]
\end{multline}
Inserting the expansion (\ref{Expand_CurMa_space}) into (\ref{GM}) and taking
the limit $Q^2 \to 0$ then
shows that terms with $\mathcal{O}(\bv{\eta}) > 1$ vanish because
$\bv{\eta}(P)$ is of order $\sqrt{Q^2}$. Concerning the first order term we
find with the special choice of the three-momentum transfer (\ref{Def_q_z})
and the definition of the rapidity (\ref{rapidity}):
\begin{equation}
\lim_{Q^2 \to 0} \frac{\bv{\eta}(P)}{\sqrt{Q^2}}
= \lim_{Q^2 \to 0} \frac{-\sqrt{Q^2}}
{2 \sqrt{M^2 + Q^2/4} \sqrt{Q^2}} \bv{e}_3
= \frac{-1}{2 M} \bv{e}_3.
\end{equation}
Therefore the static limit can safely be taken and we find for the magnetic
moment:
\begin{multline}
\mu = - \frac{3}{4 M^2}
\int\frac{d^4p_\xi}{(2\pi)^4} \int\frac{d^4p_\eta}{(2\pi)^4}
\overline\Gamma_{M, \lambda'}^{\Lambda}(\bv{p_\xi}, \bv{p_\eta}) \\[1.5ex]
\times \left[ S_F^1(p_\xi+\tf12 p_\eta) \otimes
S_F^2(-p_\xi+p_\eta) \otimes S_F^3(M-p_\eta) \right] \\[1.5ex]
\times \ii \hat{K}_3 \left[\id \otimes \id \otimes \hat{q} \gamma^+
S_F^3(M-p_\eta) \right] \Gamma_{M, \lambda}^{\Lambda}(p_\xi,p_\eta).
\end{multline}
We inserted a factor $1/2 M$ in this expression since the wavefunctions
are normalized to $2 M$ as can be seen from (\ref{NormSA}). Integration over
the relative energies can now be done after replacing the fermion propagators
by their partial fraction decomposition (\ref{PFD_prop}):
\begin{multline} \label{mu_bef_symm}
\langle \mu \rangle = \frac{3}{2 M}
\int\frac{d^3p_\xi}{(2\pi)^3} \int\frac{d^3p_\eta}{(2\pi)^3}
\overline\Gamma_{M, \lambda'}^{\Lambda}(\bv{p_\xi}, \bv{p_\eta}) \\
\left[ \frac{\Lambda^{+++}}{(M - \Omega)} + \frac{\Lambda^{---}}{(M + \Omega)}
\right] \ii F^{3+} \hat{q}_3 \\
\times \left[
\frac{\Lambda^{+++}}{(M - \Omega)} + \frac{\Lambda^{---}}{(M + \Omega)} \right]
[\gamma^0 \otimes \gamma^0 \otimes \gamma^0]
\Gamma_{M, \lambda}^{\Lambda}(\bv{p_\xi}, \bv{p_\eta}),
\end{multline}
with the tensor operator:
\begin{multline} \label{Def_Fij}
F^{ij} :=
\frac{1}{2M} \bigg\{ \frac{p_3^j}{2 \omega_3} \bigg[
\tf12 \left( \omega_1 - \omega_2 \right) \ii \tf{\partial}{\partial p_\xi^i}
+ \left( \omega_1 + \omega_2 \right) \ii \tf{\partial}{\partial p_\eta^i} \\
- \textrm{h.~c.} \bigg]
+ \frac{\Omega}{2 \omega_3}
\left( \id \otimes \id \otimes \ii \alpha^i \alpha^j \right)
+ \frac{\omega_1 + \omega_2}{2 \omega_3^2} p_3^i p_3^j \bigg\},
\end{multline}
where we also used the anticommutator $\{\gamma^0,\alpha^i\}_+=0$ and the
relation
$\Lambda_i^\pm \alpha^j = \alpha^j \Lambda_i^\mp \pm p_i^j/\omega_i$.
Note that the ``$+$''-component in the second index of $F^{ij}$ in
(\ref{mu_bef_symm}) has to be taken in the sense of (\ref{j_+}).
Before we analyze this expression further let us replace the vertex functions
in (\ref{mu_bef_symm}) by using the relations (\ref{SA_Vertex}) and
(\ref{adj_vert}) to arrive at the compact notation:
\begin{equation} \label{Mu_K}
\langle \mu \rangle = \frac{3}{2 M}
\langle \Phi_{M, \lambda'}^\Lambda | F^{3+} \hat{q}_3
| \Phi_{M, \lambda}^\Lambda \rangle
\end{equation}
Since $F^{ij}$ is a product of two vector operators it constitutes a
Cartesian tensor operator of rank $2$, which can be decomposed into
irreducible representations of the rotation group according to
(\ref{Tens_Decomp}). Just as we did when deriving the charge radius, we may
show that the contribution of certain representations vanish. The scalar
representation gives no
contribution because of the $m$-selection rule of the Wigner-Eckart theorem,
which states that
\begin{equation}
\langle j_1, m_1 | F^{[k]}_q | j_2, m_2 \rangle = 0\quad \textrm{unless}\quad
m_1 - m_2 = q.
\end{equation}
In our case $m_1 = \tf12$, $m_2 = -\tf12$ and $q = 0$. The
spin $2$ representation vanishes because of the triangularity relation:
\begin{equation}
\langle \tf12, \tf12 | T^{[2]}_q | \tf12, -\tf12 \rangle = 0.
\end{equation}
We are thus left with the antisymmetric
representation belonging to spin $1$ which we may write as a vector product:
\begin{equation} \label{F_spherical}
{F^{3+}}^{[1]}
= \left( F^{31} + \ii F^{32} \right)^{[1]}
= \tf{\ii}{\sqrt{2}} \sum_{j,k=1}^3 \epsilon_{+jk} F^{jk}
\end{equation}
where it is understood to take the spherical ``$+1$''-component of the vector
product as defined in (\ref{vector_operator}). Note that since
$F^{ij}$ is contracted with the skew tensor $\epsilon_{ijk}$, the last term in
(\ref{Def_Fij}) that is proportional to $p_3^i p_3^j$ vanishes.
Inserting (\ref{F_spherical}) back into (\ref{Mu_K}) and choosing the spin
projections $\lambda'=\tf12$ and $\lambda=-\tf12$ then yields:
\begin{equation} \label{Mu_K_+1}
\langle \mu \rangle = \frac{3}{2 M}
\langle \Phi_{M, 1/2}^\Lambda |
\tf{1}{\sqrt{2}} \sum_{j,k=1}^3 \epsilon_{+jk} F^{jk} \hat{q}_3
| \Phi_{M, -1/2}^\Lambda \rangle.
\end{equation}
By using the Wigner-Eckart theorem once again
we remove the spin-flip and turn the expression in an expectation value:
\begin{equation} \label{mu_final}
\langle \mu \rangle = - \frac{3}{2 M}
\langle \Phi_{M, 1/2}^\Lambda |
\sum_{j,k=1}^3 \epsilon_{3jk} F^{jk} \hat{q}_3
| \Phi_{M, 1/2}^\Lambda \rangle,
\end{equation}

To simplify (\ref{Def_Fij}) further we replace the relative coordinates by
one-particle coordinates:
\begin{multline}
F^{ij} := \frac{1}{2M} \Bigg\{ \frac12 \left[ - \frac{\Omega}{\omega_3}
p_3^j \bigg( \ii \tf{\partial}{\partial p_3^i}
- \frac1\Omega \sum_{\alpha=1}^3 \omega_\alpha \ii
\tf{\partial}{\partial p_\alpha^i} \right) \\
- \textrm{h.~c.} \bigg]
+ \frac{\Omega}{2 \omega_3}
\left( \id \otimes \id \otimes \ii \alpha^i \alpha^j \right) \Bigg\}.
\end{multline}
Since in the expectation value (\ref{mu_final}) $F^{jk}$ is contracted with
the skew symmetric tensor $\epsilon_{ijk}$ it is suggested to define:
\begin{equation} \label{Def_LR}
\hat{L}_{R\alpha}^i := \epsilon_{ijk} p_\alpha^k \left(
\ii \tf{\partial}{\partial p_\alpha^j} - \hat{R}^j \right).
\end{equation}
$\hat{L}_{R\alpha}^i$ is obviously the total angular momentum of the
three-quark system with the correct center of mass motion removed.
Furthermore we identify the spin operator $\bv{S} = \tf12 \bv{\Sigma}$ in the
following contraction:
\begin{equation}
\sum_{j,k=1}^3 \epsilon_{ijk} \alpha^j \alpha^k
= 2 \ii
\left( \begin{array}{ccc} \sigma_i & \id \\ \id & \sigma_i \end{array} \right)
= 2 \ii \Sigma^i.
\end{equation}
We then have:
\begin{equation} \label{eps_Fij}
\sum_{j,k=1}^3 \epsilon_{ijk} F^{jk}
= -\frac{\Omega}{4 M \omega_3} \left(
\hat{L}_{Ri}^3 + \id \otimes \id \otimes \Sigma^i
+ \textrm{h.~c.} \right).
\end{equation}
This expression is still not symmetric in the three fermions, so
in the final step we symmetrize over the three fermions in the same way as we
did already when deriving the charge radius:
\begin{equation} \label{MaMo_ExpValue}
\langle \mu \rangle = 
\frac{\langle \Phi_M^\Lambda | \hat{\mu} | \Phi_M^\Lambda \rangle}
{\langle \Phi_M^\Lambda | \Phi_M^\Lambda \rangle},
\end{equation}
where we defined the magnetic moment operator $\hat{\mu}$ which follows from
symmetrizing (\ref{eps_Fij}):
\begin{equation} \label{mu_op}
\hat{\mu} = \frac12 \left[ \frac{\Omega}{M} \sum_{\alpha=1}^3
\frac{\hat{q}_\alpha}{2 \omega_\alpha}
\left( \hat{L}_{R\alpha}^3 + 2 S_\alpha^3 \right)
+ \textrm{h.~c.} \right].
\end{equation}
with the one-particle spin operators:
\begin{align}
\bv{S}_1 &:= \bv{\Sigma}/2 \otimes \id \otimes \id  \nonumber \\[1.5ex]
\bv{S}_2 &:= \id \otimes \bv{\Sigma}/2 \otimes \id \\[1.5ex]
\bv{S}_3 &:= \id \otimes \id \otimes \bv{\Sigma}/2. \nonumber
\end{align}
\subsection{Interpretation}
As has already been shown in the interpretation of the charge radius, the term
(\ref{pos_cms}) corresponds to the position of particle $\alpha$ as measured
from the center of mass of the system. One is thus naturally led to interpret
$\bv{\hat{L}_{R\alpha}}$ defined in (\ref{Def_LR}) as the angular momentum
(operator) observed from the relativistic center of mass. As already mentioned
$\bv{S_1}$, $\bv{S_2}$ and $\bv{S_3}$ are one-particle spin operators. We
therefore conclude that
the magnetic moment of the system can be decomposed in contributions of the
fermion angular momenta and their spins:
\begin{equation} \label{mu_decomp}
\langle \mu \rangle = \langle \mu_L \rangle + 2 \langle \mu_S \rangle,
\end{equation}
with $\langle \mu_L \rangle$ being the contribution of the angular momenta of
the three fermions:
\begin{multline} \label{ang_mom_contr}
\langle \mu_L \rangle :=
\frac{1}{\langle \Phi_M^\Lambda | \phi_M^\Lambda \rangle}
\langle \Phi_{M}^\Lambda |\,
\frac12 \bigg( \frac{\Omega}{M} \sum_{\alpha=1}^3
\frac{\hat{q}_\alpha}{2 \omega_\alpha} \hat{L}_{R\alpha}^3 \\
+ \textrm{h.~c.} \bigg)\,
| \Phi_{M}^\Lambda \rangle
\end{multline}
and $\langle \mu_S \rangle$ the contribution of the fermion spins:
\begin{equation} \label{spin_contr}
\langle \mu_S \rangle :=
\frac{1}{\langle \Phi_M^\Lambda | \phi_M^\Lambda \rangle}
\langle \Phi_{M}^\Lambda |\,
\frac{\Omega}{M} \sum_{\alpha=1}^3 S_\alpha^3\,
| \Phi_{M}^\Lambda \rangle.
\end{equation}
Such a decomposition into spin and angular momentum contributions is not
possible when extracting the magnetic moment from a form factor. It is thus
another benefit of the approach to static properties presented in this work.
In (\ref{mu_op}) we discover the same relativistic weight factor $\Omega/M$
as has already been found in the charge radius. When taking the nonrelativistic
limit, the operator $\hat{\mu}$ (\ref{mu_op}) becomes:
\begin{multline} \label{mu_op_nr}
\hat{\mu}_{\textrm{n.r.}} =
\sum_{\alpha=1}^3 \frac{\hat{q}_\alpha}{2 m_\alpha}
\epsilon_{3jk} p_\alpha^j \left( \ii \tf{\partial}{\partial p_\alpha^k}
- \frac1M \sum_{\beta=1}^3 m_\beta \ii
\tf{\partial}{\partial p_\beta^i} \right) \\
+ 2 \sum_{\alpha=1}^3 \frac{\hat{q}_\alpha}{2 m_\alpha} S_\alpha^3.
\end{multline}
Except for the center of mass correction this expression is well known.
We are thus led to conclude that we have found the relativistic generalization
of the nonrelativistic magnetic moment operator.

Although the expression (\ref{Def_MaMo_FF}) we started from holds only for
spin $1/2$ states, one can, as in the case of the charge radius, generalize the
formalism to arbitrary spins and the result is not different from the
spin $1/2$ case.
\section{Extension to higher moments}
The formalism presented so far paves the way for the calculation of higher
moments as well -- an issue that we briefly want to touch upon in this section.
We take the electric form factor as an example and work accordingly with the
``time''-component of the current matrix element (\ref{CurMa_time}).
An arbitrary moment $\langle m \rangle$ of a charge distribution is then
given in general by:
\begin{equation} \label{ext_start}
\langle m \rangle =
\sum_{i_1, i_2,\dots,i_n=1}^3 O_{i_1 i_2 \dots i_n}
\int \dd^3x\, x^{i_1} x^{i_2} \cdots x^{i_n} \rho(\bv{x}),
\end{equation}
where $O_{i_1,i_2,\dots,i_n}$ is a tensor of rank $n$, which depends on the
moment to be computed. For example for the charge radius, considered so far,
$O$ is simply:
\begin{equation} \label{Def_O_cr}
O_{i_1 i_2} = \frac{1}{Q} \delta_{i_1 i_2}.
\end{equation}
By similar steps leading from eq. (\ref{BasicChaRadDef}) to
eq. (\ref{ChaRadCurDis}) we get:
\begin{multline} \label{m_breit}
\langle m \rangle =
\frac{1}{2M} \left( \frac{-\ii}{2} \right)^n
\sum_{i_1, i_2,\dots,i_n=1}^3 O_{i_1 i_2 \dots i_n} \\
\times \frac{\partial}{\partial P^{i_1}} \frac{\partial}{\partial P^{i_2}}
\cdots \frac{\partial}{\partial P^{i_n}}
\langle \mathcal{P} P | j^0(0) | P \rangle \bigg|_{\bv{P}=0}.
\end{multline}
The current matrix element appearing here is defined in eq. (\ref{CurMa_time}).
As before its $P$-dependent part is given by an exponential of
infinitesimal boosts as in eq. (\ref{Boost}). Because
\begin{equation}
\lim_{\bv{P} \to 0} \eta(P)=0\quad \mathrm{and}\quad 
\frac{\partial}{\partial P^i} \eta^j(P) \bigg|_{\bv{P}=0}
= \frac{\delta_{ij}}{M},
\end{equation}
we find:
\begin{multline}
\frac{\partial}{\partial P^{i_1}} \frac{\partial}{\partial P^{i_2}}
\cdots \frac{\partial}{\partial P^{i_n}}
\exp(-2 \ii \bv{\eta}(P) \cdot \bv{\hat{K}}) \bigg|_{\bv{P}=0} \\
= \frac{(-2 \ii)^n}{M^n} \hat{K}^{i_1} \hat{K}^{i_2} \cdots \hat{K}^{i_n}
\end{multline}
Note that to every $x^i$ from our starting equation (\ref{ext_start})
corresponds now a boost generator $\hat{K}^i$. Using this result we get from
eq.
(\ref{m_breit}):
\begin{multline}
\langle m \rangle =
\frac{1}{2M} \sum_{i_1, i_2,\dots,i_n=1}^3 O_{i_1 i_2 \dots i_n}\, \\
\times (-3) \int\frac{d^4p_\xi}{(2\pi)^4} \int\frac{d^4p_\eta}{(2\pi)^4}
\overline\Gamma_{M}^{\Lambda}(\bv{p_\xi}, \bv{p_\eta}) \\
\times \left[ S_F^1(p_\xi+\tf12 p_\eta) \otimes
S_F^2(-p_\xi+p_\eta) \otimes S_F^3(M-p_\eta) \right] \\
\times \frac{1}{M^n} \hat{K}^{i_1} \hat{K}^{i_2} \cdots \hat{K}^{i_n} \\
\times \left[\id \otimes \id \otimes  \gamma^0 \hat{q}
S_F^3(M-p_\eta) \right]
\Gamma_{M}^{\Lambda}(p_\xi, p_\eta).
\end{multline}
Integrating out the dependence on the relative energies after replacing the
propagators according to eq. (\ref{PFD_prop}) then results in:
\begin{multline} \label{Arb_mom_final}
\langle m \rangle =
\frac{3}{\langle \Phi_M^\Lambda | \Phi_M^\Lambda \rangle} \\
\times \sum_{i_1, i_2,\dots,i_n=1}^3 O_{i_1 i_2 \dots i_n}\,
\langle \Phi_M^\Lambda |
\frac{1}{M^n} \hat{K}'_{i_1}  \hat{K}'_{i_2} \dots  \hat{K}'_{i_n}
\hat{q}_3 | \Phi_M^\Lambda \rangle \\
+\, \textrm{off-diagonal matrix elements}
\end{multline}
where $\hat{K}'_i$ is defined in eq. (\ref{K'^i}).
For $n>2$ we also find terms involving matrix elements between different energy
components of the vertex function, i.~e. between the subspaces of purely
positive and negative energy components (denoted
``off-diagonal matrix elements'' in eq. (\ref{Arb_mom_final})). Unfortunately
these terms cannot be
expressed in a generic way and have to be calculated explicitly for the moment
under consideration. One might however expect that these additional
contributions are in fact small; first because the negative energy components
correspond to the ``small'' components of the Dirac equation and thus vanish in
the non-relativistic limit and second because both energy subspaces are
orthogonal. Note that although the first term of eq.
(\ref{Arb_mom_final}) also involves matrix elements between different energy
subspaces of the Salpeter amplitudes, one can show that these do in fact
vanish.

Finally we may symmetrize the expectation value in eq. (\ref{Arb_mom_final})
over the three fermions to obtain:
\begin{multline} \label{Arb_mom_final_symm}
\langle m \rangle =
\frac{1}{\langle \Phi_M^\Lambda | \Phi_M^\Lambda \rangle}
\sum_{i_1, i_2,\dots,i_n=1}^3 O_{i_1 i_2 \dots i_n}\, \\
\times \langle \Phi_M^\Lambda |
\sum_{\alpha=1}^3 \hat{K}''_{i_1\, \alpha}  \hat{K}''_{i_2\, \alpha} \dots
\hat{K}''_{i_n\, \alpha}
\hat{q}_\alpha | \Phi_M^\Lambda \rangle \\
+\, \textrm{off-diagonal matrix elements},
\end{multline}
where $\hat{K}''_{i\, \alpha}$ is defined as:
\begin{equation}
\hat{K}''_{i\, \alpha} = \frac{1}{2}
\left[ \frac{\Omega}{M} \left(\ii \frac{\partial}{\partial p^i_\alpha}
- \bv{\hat{R}} \right) + \mathrm{h.~c.} \right].
\end{equation}
If e.~g. we insert eq. (\ref{Def_O_cr}) into the final result
(\ref{Arb_mom_final_symm}) we instantly obtain the charge radius expression
(\ref{CharRad_boost_final}). What has been said about its interpretation also
applies to eq. (\ref{Arb_mom_final_symm}) in its general form.

In this sense a generalization of the formalism presented in this work to
arbitrary moments is possible, although those discussed in detail, namely
the charge radius and the magnetic moment are by far the most interesting,
having in addition the soundest empirical basis.
\section{Application to static properties of baryons}
We would like to illustrate the relevance of the preceeding theoretical
considerations by applying them to an existing physical model. In refs.
\cite{Loring:2001kv,Loring:2001kx,Loring:2001ky} a relativistic covariant quark
model
for baryons is treated, based on assumptions which also entered the work at
hand, i.~e. instantaneous interaction kernels and free fermion propagators
corresponding to effective fermion masses. The model successfully describes
mass spectra of strange and non-strange baryons up to the highest orbital
and radial excitations employing a linear confinement potential and a residual
interaction based on an effective instanton force. The seven parameters
entering the model are fixed by a fit to the best established resonances.
We use the wavefunctions, i.~e. Salpeter amplitudes, that have been obtained
by solving the Salpeter equation within this model to compute the expectation
values of the charge radius and magnetic moment operator derived in this work.
Since no further parameters are introduced the results are genuine predictions.
\subsection{Nucleon charge radii}
The proton charge radius that we obtain by computation of the expectation
value (\ref{CharRad_boost_final}) amounts to
\begin{equation}
\sqrt{\langle r^2 \rangle_\mathrm{proton}} = 0.86\, \mathrm{fm},
\end{equation}
in excellent agreement with the experimental value of \\
$0.87 \pm 0.008\, \mathrm{fm}$ from ref. \cite{Eidelman}.
The mean square charge radius of the neutron however results in
\begin{equation}
\langle r^2 \rangle_\mathrm{neutron} = - 0.206\, \mathrm{fm}^2
\end{equation}
and overestimates the empirical number of\\
$-0.1161 \pm 0.0022\, \mathrm{fm}^2$ from ref. \cite{Eidelman} by $77\, \%$.
Within the same model the authors of ref. \cite{Merten:2002nz} have calculated
the neutron electric form factor and extracted a mean square charge radius of
$-0.11\, \mathrm{fm}^2$ from it. The procedure however was numerically
erroneous and a reanalysis, improving the numerical precision, resulted in a
radius that is indeed compatible with our result.
\subsection{Baryon octet magnetic moments}
In the same model we have computed the nucleon magnetic moments using our
formula (\ref{MaMo_ExpValue}). For the proton we find a magnetic moment of
\begin{equation}
\langle \mu \rangle_\mathrm{proton} = 2.77\, \mu_N
\end{equation}
in perfect agreement with the empirical value of $2.793\, \mu_N$.
The magnitude of the neutron magnetic moment
\begin{equation}
\langle \mu \rangle_\mathrm{neutron} = -1.71\, \mu_N
\end{equation}
is rather small if compared to the experimental magnetic moment
of $-1.913\, \mu_N$.

In addition to the nucleon magnetic moments we have also calculated those of
the strange octet baryons because they are experimentally well covered.
Table \ref{MaMo_hyperons} compares our results to the empirical values.
\begin{table}[h]
\begin{center}
\renewcommand{\arraystretch}{1.5}
\begin{tabular}{c||r|r} \hline
hyperon & experiment \cite{Eidelman}& this calculation \\
 & [$\mu/\mu^{}_N$] & [$\mu/\mu^{}_N$] \\ \hline
$\Lambda$ & $-0.613 \pm 0.004$ & $-0.61$ \\
$\Sigma^+$ & $2.458 \pm 0.01$ & $2.51$ \\
$\Sigma^0$ & $-$ & $0.75$ \\
$\Sigma^-$ & $-1.16 \pm 0.025$ & $-1.02$ \\
$\Xi^0$ & $-1.25 \pm 0.014$ & $-1.33$ \\
$\Xi^-$ & $-0.6507 \pm 0.0025$ & $-0.56$ \\ \hline
\end{tabular}
\caption{Hyperon magnetic moments compared to the empirical values.}
\label{MaMo_hyperons}
\renewcommand{\arraystretch}{1}
\end{center}
\end{table}
The results are in excellent agreement with experiment. The largest deviation
of $14\, \%$ is seen with the $\Xi^-$-magnetic moment.

Since efforts are being made to measure also magnetic moments of excited
nucleon states like the $\mathrm{S}^{11}(1535)$ as mentioned in ref.
\cite{Kotulla:2002cg}, we contribute some selected predictions here. The
magnetic
moments of the nucleon Roper resonance ($\mathrm{P}^{11}(1440)$) and the lowest
lying state with total spin $1/2$ and negative parity ($\mathrm{S}^{11}(1535)$)
are shown in table \ref{MaMo_excited}.
\begin{table}[h]
\begin{center}
\renewcommand{\arraystretch}{1.5}
\begin{tabular}{c||r|r} \hline
nucleon resonance & $I_3$ & magnetic moment \\
 & & [$\mu/\mu^{}_N$] \\ \hline
$\mathrm{P}^{11}(1440)$ & $1/2$ & $1.55$ \\
 & $-1/2$ & $-0.98$ \\ \hline
$\mathrm{S}^{11}(1535)$ & $1/2$ & $0.37$ \\
 & $-1/2$ & $-0.1$ \\ \hline
\end{tabular}
\caption{Prediction of magnetic moments of selected excited nucleon states.
$I_3$ means third isospin component.}
\label{MaMo_excited}
\renewcommand{\arraystretch}{1}
\end{center}
\end{table}
The formalism allows the computation of magnetic moments of baryons with
arbitrary spins and their radial excitations which will be the subject of
a subsequent publication. The same is true of course for the charge radius.

As has already been indicated, the magnetic moment may be decomposed in spin
and angular momentum contributions according to eq. \ref{mu_decomp}.
This decomposition enables us to carry out a numerical analysis of the
magnitudes of both spin and angular momentum contributions. Note that such a
study is not possible
by relying on form factor calculations because there only the total magnitude
of the magnetic moment can be extracted. Table \ref{decomposition} lists the
contributions of spin and angular momentum to the magnetic moments of proton
and neutron.
\begin{table}[h]
\begin{center}
\renewcommand{\arraystretch}{1.5}
\begin{tabular}{c||r|r||r|r} \hline
& $2\langle \mu_S \rangle$ &
$\frac{2\langle \mu_S \rangle}{\langle \mu\rangle}$ &
$\langle \mu_L \rangle$ &
$\frac{\langle \mu_L \rangle}{\langle \mu\rangle}$ \\
& $[\mu/\mu_N]$ & $[\%]$ & $[\mu/\mu_N]$ & $[\%]$ \\ \hline
proton & $2.53$ & $91$ & $0.24$ & $9$ \\
neutron & $-1.59$ & $93$ & $-0.12$ & $7$ \\ \hline
\end{tabular}
\caption{Contributions of quark spins ($2 \langle \mu_S \rangle$) and angular
momentum ($\langle \mu_L \rangle$) to the net magnetic moments of proton and
neutron.}
\label{decomposition}
\renewcommand{\arraystretch}{1}
\end{center}
\end{table}
The analysis shows, that the contribution of the quark spins exceeds the
contribution of the quark angular momenta by far. One can state that a good
$90\%$ of the magnetic moment is coming from quark spins which is due to the
fact that the
quarks are dominantly in a relative $S$-wave. This result also explains in part
the success of the nonrelativistic quark model in predicting the magnetic
moments. Our analysis shows that by neglecting the angular motion of the quarks
by assuming that the quarks are in a relative $S$-wave, the induced error is in
the percent region. We should mention that for the $\mathrm{S}^{11}(1535)$
the absolute value of the spin contribution is only a quarter of the angular
momentum contribution and opposite in sign. Since this resonance is dominantly
a $P$-wave the spin has to be aligned antiparallel to the angular momentum
to result in a state with total spin $1/2$.
The preceeding discussion is however only true if we work with a constituent
quark mass of 330 MeV.

\begin{figure} \label{evolution_nucleon}
\begin{center}
\includegraphics[bb=116 623 352 839,clip,width=0.8\linewidth]{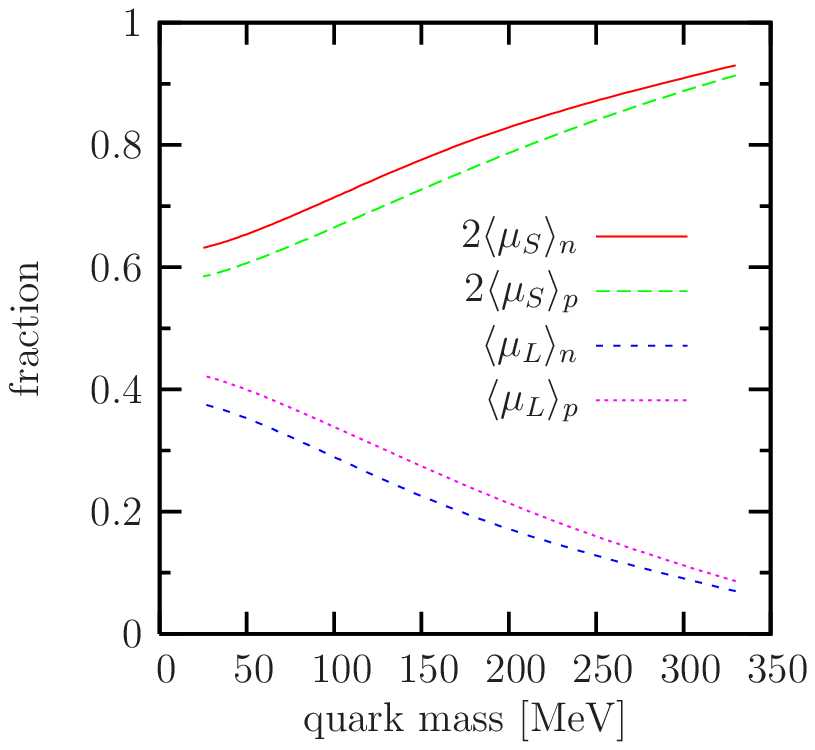}
\caption{Fraction of the total magnetic moment carried by quark spin and
angular momentum respectively of proton and neutron respectively as a function
of the quark mass.}
\end{center}
\end{figure}

We may however carry this analysis further by studying the evolution of the
spin/angular momentum distribution with smaller quark masses. Note that the
quark
model described in refs. \cite{Loring:2001kv,Loring:2001kx,Loring:2001ky}
assumes
isospin symmetry between up- and down-quark and thus there is only one mass
parameter for the nucleon. At different magnitudes of this mass parameter we
have now fitted the remaining six
parameters of the model to the baryon spectra. We might of course not expect to
reproduce the spectra as well as with the original value of
$330\, \mathrm{MeV}$ but at least we were able to keep the ground states i.~e.
the nucleon and the $\Delta$-particle at the empirical values. We achieved a
quark mass as small as
$25\, \mathrm{MeV}$ before numerical restrictions impeded us to go any
further. Figure \ref{evolution_nucleon} shows the effect on the spin and
angular momentum contribution to the magnetic moment of proton and neutron.
We see an almost linear decrease of the spin contribution from its original
value of a good $90\, \%$ at $330\, \mathrm{MeV}$ to roughly $60\, \%$ at
$25\, \mathrm{MeV}$. At the same time the angular momentum contribution gains
in magnitude correspondingly to roughly $40\, \%$.
Although one looses the concept of constituent quarks at such small quark
masses, the analysis nevertheless shows that when approaching the chiral limit,
spin and angular momentum contribution to the magnetic moment become of the
same order of magnitude.
\section{Conclusion}
We have shown how the charge radius and the magnetic moment of a bound
three-fermion system with instantaneous interactions can be formulated as
expectation values with respect to Salpeter amplitudes. The corresponding
operators turned out to be natural relativistic generalizations of their
non-relativistic counterparts. We also indicated how the formalism may be
extended to higher moments as well. A first application of the formalism to a
relativistic quark model for baryons with instantaneous interactions described
in refs.
\cite{Loring:2001kv,Loring:2001kx,Loring:2001ky} results in a good description
of the
nucleon charge radii and baryon octet magnetic moments except for the neutron
radius. Predictions have been made for the magnetic moments of the
$\mathrm{P}^{11}(1440)$ and $\mathrm{S}^{11}(1535)$ resonances.
We found in addition an interesting dependence of the nucleon magnetic moments
on the quark masses. In
particular, when constituent quark masses are decreased to values almost as
small as current masses, the
spin contributions become equal in magnitude in comparison to the contributions
of internal angular momenta. Static observables of systems with spins other
than $1/2$ like e.~g. the baryon decuplet will be studied in the future.
\section*{Acknowledgments}
Financial support from the Deutsche Forschungsgemeinschaft by the
SFB/Transregio 16 ``Subnuclear Structure of Matter'' and from the European
Community-Research Infrastructure Activity under the FP6 "Structuring the
European Research Area" programme (HadronPhysics, contract number
RII3-CT-2004-506078) is gratefully acknowledged.
\end{document}